# Mathematical Modeling of Flash Boiling Phenomena in Superheated Sprays at low Degree of Superheat using Dirichlet Hyperboloids

B. Bhatia,[1] A. De,[1,2,*] and E. Gutheil[2]

[1] *Department of Aerospace Engineering, Indian Institute of Technology Kanpur, 208016, Kanpur, India.*

[2] *Interdisciplinary Center for Scientific Computing (IWR), Heidelberg University, 69120, Heidelberg, Germany.*

The depressurization of the surrounding chamber or superheating of the injected liquid is responsible for the flashing of the sprays, which promotes micro-explosion of many bubbles near free-surface and thereby leading to primary atomization of sprays. In this work, we propose a mathematical model, and the Dirichlet hyperboloids are used to explain the micro-explosion process while the external flashing phenomenon is observed in superheated liquid jets. The developed mathematical model is implemented in the Lagrangian framework to study the spray structure, and the results of numerical simulations are found to be in good agreement with the experimental results from the literature. In the onset region of a fully flashing regime, the bell-shaped spray structure becomes eminent due to increased drag on the high radial-velocity ejected droplets. However, at a lower degree of superheat, the droplet size is found to increase with a decrease in ambient pressure. Whereas the opposite trend is observed at a higher degree of superheat.

## 1. Introduction

Though there have been many studies on flash-boiling phenomenon lately, the problem is quite several decades old and equally challenging. The flashing of liquids poses a grave danger of accidental releases and explosions in chemical, cryogenic, or nuclear industries. On the other hand, the flashing of liquid jets at a higher degree of superheat is advantageous in fine atomization and subsequently mixing of fuel with oxidizer before combustion, hence necessary in the engine design. Some notable characteristic features of flashing jets include high evaporation rate, higher spread angle, low liquid penetration length, and smaller droplet size (Kim et al. 1999).

Oza (1984) and Oza and Sinnamom (1983) segregated the flashing jets into "internally flashing" and "externally flashing" regimes. The two regimes differ from each other concerning the location of the bubble inception region and whether the bubble grows within or outside the nozzle. The bubble growth or liquid evaporation inside the nozzle yields liquid jet breakup directly at the nozzle exit. They suggested that a high degree of superheating will make internally flashing regime dominant.

---

* Corresponding author. E-mail addresses: ashoke@iitk.ac.in

The results of the experiments conducted by Reitz (1990) reported that a high degree of superheat of the injected liquid does not necessarily show internal flashing and possesses an intact liquid core at the nozzle exit, within a surrounding spray plume. But, Park and Lee (1994) found out that the internal flow pattern before discharge from the nozzle governed the spray characteristics outside the nozzle. They suggested a length-to-diameter (L/D) ratio of seven for a nozzle to have a transition between the two modes. Günther and Wirth (2013), in their recent experimental study, emphasized the critical role of the nozzle material in determining the initiation of internal flashing at a high degree of superheat. They also measured the acoustic signal or the "pulsation frequencies" generated by the bubble bursts and the subsequent vapor formation at a high degree of superheat, for glass and steel nozzles. As concluded by them, it is a higher roughness of steel that initiated internal flashing at a lower superheat temperature compared to the glass nozzles. Hence, with a smooth nozzle or at a lower degree of superheat, externally flashing sprays are dominant.

Further, study by Wu et al. (2017) exhibits the dominant role of the superheat degree over ambient pressure on the amount of internal flashing that happens within the nozzle. Thus, a study involving the varying ambient pressure will be suitable for the externally flashing jets (Kamoun et al. 2010; Lamanna et al. 2014; Kamoun et al. 2014). The externally flashing sprays being less dependent on the nozzle internal geometry, material, and its surface finish, are not limited to a few specific cases, and hence are more intriguing for a study. The study by Lecourt et al. (2009) provided the spray characteristics at near-vacuum conditions for different fuels and injector designs. They witnessed the external flashing of sprays at a high fuel flow rate for a single orifice injector.

There has been a considerable effort in modeling or arriving at a correlation for flash-boiling owing to the specific nozzle design and related complexities associated with the internally flashing sprays. Some essential aspects of sprays still require modeling: the transition condition – from mechanical to flash-boiling; the spray structure – droplet size, spread-angle, and the vaporization rate. Brown and York (1962) proposed a critical Weber number corresponding to a superheat degree, above which flash atomization occurs. Kitamura et al. (1986) found a critical superheat for the flashing breakup, expressed as the Jacob Number (Ja), corrected with a correction factor ($\varphi$) as a function of jet Weber number. Both these later models are limited to high Weber number jets. Cleary et al. (2007) and Witlox et al. (2007) developed an empirical model to predict the droplet size and its distribution. It also defines a critical Weber number and conditions of transition between flashing and non-flashing jets. The reported results were found to be over-predicted in some cases. More accurate condition of transition and the onset of flashing regime were redefined by Lamanna et al. (2009). They also studied the spray structure and proposed a universal correlation for the spread angle. However, the empirical correlation is valid in a completely flashing regime and limited to a particular downstream distance from the nozzle exit.

For externally flashing sprays with a lower degree of superheat, the presence of a liquid core is an essential consideration in an empirical or mathematical model. Lienhard and Day (1970) established the condition and average length of liquid core breakup. According to a model proposed by Sher and Elata (1977), bubbles grow until they touch each other, and the liquid core flashes into an aerosol. Based on the experimental observation of Suma and Koizumi (1977), the liquid core breaks-up when the void fraction reaches a particular constant value in the models of Adachi et al. (1997) and Senda et al. (1994).

As mentioned above, several studies predict the transition range for the flashing break-up. On the other hand, a scant number of models are available for the liquid core break-up due to flash-boiling, especially at a low superheat regime of heterogeneous nucleation. The absence of a complete physical model in the existing literature that predicts not only the spray structure, but also the liquid core breakup, droplet size, and velocities, is the motivation for the present work. The work of Khan et al. (2017) and Rodrigues et al. (2015) strongly suggests the significance of the spray structure in the internal combustion engines and other reactive systems. Hence, the current study aims at describing the underlying phenomena involved in the liquid jet-core break-up and developing a mathematical model that predicts the liquid jet-core breakup/ bubble micro-explosions, the resulting droplet size and velocity, and the near-nozzle spray structure.

*1.1. Flash-Boiling*

A superheated liquid at high pressure, when exposed to low pressure, may remain in the metastable liquid state, provided the degree of superheat is less than the thermal kinetic limit (as defined by spinodal curve), and the probable sites for heterogeneous nucleation are absent. Once the spinodal superheat limit is crossed, the homogeneous nucleation becomes dominant, resulting in higher bubble nuclei formation (Skripov 1974). At a higher degree of superheat, an increased amount of nucleation causes the faster transformation of the metastable liquid to two-phase flow. This feature of flash-boiling is evident in spray studies, where a superheated liquid at higher pressure is injected into a low-pressure chamber (Avdeev 2016), or the studies of depressurization of superheated liquid-filled vessels (Reinke and Yadigaroglu 2001). The study of superheated sprays is the main focus of this work.

As mentioned above, the bubbles predominantly nucleate heterogeneously near the nozzle exit at superheat temperatures lower than the spinodal limit. In this range of degree of superheat, the liquid jet-core exhibits a noticeable feature of the flashing spray (Avdeev 2016). This range coincides with the “transition” regime mentioned in some of the literature (Kamoun et al. 2010; Kamoun et al. 2014; Cleary et al. 2007). The nucleated bubbles of a size larger than a critical size grow as they are convected downstream in the liquid jet-core. Once present near the free surface of the liquid jet-core, the bubbles burst, ejecting smaller droplets along with the liquid-vapor into the surrounding. These ejected droplets change the spray structure and increase the spread angle. The spray structure is nearly conical in shape, with a spread angle roughly the same in the vicinity of the nozzle. This spray structure is different from the one witnessed at a superheat temperature higher than the spinodal limit. In the latter case, the spray structure is a bell-shaped with continuously varying spread angle as one traverses downstream of the nozzle exit. The radial expansion of small liquid droplets and vapor, right at the nozzle exit, is the main reason for the bell-shaped spray structure. There are various attempts to predict this latter behavior of the sprays at superheat temperatures higher than the spinodal limit (Kamoun et al. 2010; Lamanna et al. 2014; Avdeev 2016). On the contrary, to the best of the authors' knowledge, there exists no complete model which physically explains the liquid jet breakup, the resulting spray structure, ejected particle size, and velocities at a lower degree of superheat.

After traversing through the liquid jet-core at a lower superheat, the bubbles near the free surface undergo *micro-explosions* or bubble bursts. Hence the complete journey of a bubble from nucleation to the burst can be divided as 1) nucleation, 2) bubble growth, and 3) bubble-burst or micro explosion. As discussed in the last stage, which is very significant and controls several characteristics of flashing sprays, while the first two stages are widely studied and documented in a large number of previous literature.

### *1.1.1 Bubble Nucleation*

It is a crucial factor in deciding the amount of vaporization of the liquid phase. It may be heterogeneous or homogeneous based on the degree of superheat. At a high degree of superheat, more than the spinodal limit, homogeneous nucleation occurs (Skripov et al. 1974; Liao and Lucas 2017). A notable study on the homogeneous nucleation is by Blander and Katz (1975), which revolves around the work done by individual molecules to form a surface of a critical-sized nucleus. The nucleation rate is a product of the rate of formation of such nuclei, at a steady-state, times the concentration of nuclei where a Boltzmann distribution approximates the concentration. As pointed out by them (Blander and Katz 1975), the homogeneous nucleation takes place when the contact angle between the volatile liquid and the surfaces with which it is in contact is zero and results in explosive boiling. Many later studies have discussed the limitations of this theory (Oxtoby and Evans 1988; Talanquer and Oxtoby 1995; Oxtoby 1998) and proposed various modifications (Lubetkin and Blackwell 1988; Delale et al. 2003). Citing the deviation of this nucleation theory in highly superheated conditions, Lamanna et al. (2014) expressed the concentration of such bubble nuclei to depend on the logarithm of square of pressure. Both the studies (Kamoun et al. 2010 and Lamanna et al. 2014) attributed the flash boiling behavior to the nucleation rate of the bubbles.

The heterogeneous nucleation starts much earlier at a lower degree of superheat and vis-à-vis homogeneous nucleation, and it leads to less violent boiling. Unlike homogeneous nucleation, heterogeneous nucleation occurs at an interface between the volatile liquid and some other phase in contact, with which the contact angle is greater than zero (Blander and Katz 1975). The other phases may be a smooth/rough rigid wall or a liquid-liquid interface. In a review by Liao and Lucas (2017), two types of heterogeneous nucleation mechanisms are present in a flashing process – bulk heterogeneous and wall nucleation. The bulk heterogeneous mechanism utilizes two methods – modified classical homogeneous nucleation theory (Alamgir and Lienhard 1981; Valero and Parra 2002) and, presumed probability density function (PDF) approach (Kumzerova and Schimdt 2003). The cavities, cracks, and other irregularities in the wall are the sites for the wall nucleation process, which is developed for both smooth walls (Blander and Katz 1975; Alamgir and Lienhard 1981; Rohatgi Reshotko 1975) and rough walls (Riznic et al. 1987; Kolev 2005; Jones and Shin 1984Blinkov at al. 1993).

Wall nucleation models for smooth walls are generally limited to vessels with degassed, unscratched glass surfaces, and not applicable to crystalline solids with the ever-present irregularities (Blander and Katz 1975). In most practical cases, the nozzles are made of metals like steel, which encourages the use of wall nucleation models for rough surfaces. One of the few critical wall nucleation models, the Riznic model (Riznic et al. 1987), uses a modified form of relation developed by Kocamustafaogullari and Ishii (1983) for nucleation sites in boiling. Similarly, Kolev (2005) developed a model for the nucleation rate taking into account the heat transfer between the boundary layer and the bulk liquid. Jones and Shin (1984) used an empirical correlation for the activated nucleation site and estimated the departure frequency and size as:

$$J_{het,W} = N_a \cdot f \tag{1}$$

where,

$$N_a = 0.25 \times 10^{-7} \frac{R_d^2}{R_{cr}^4}, \text{ and } f = 10^4 \cdot (T_l - T_{sat})^3 .$$

Here, $J_{het,W}$ is the number of bubbles nucleated, $N_a$ is the number of activated nucleation sites, $f$ is the nucleation departure frequency. $R_d$ is a departure radius of the bubble, while $R_{cr}$ is the critical radius of the bubble $(= 2\sigma/(P_{sat} - P_l))$. In this study, the pressure of liquid surrounding the bubble $(P_l)$ is same as the ambient chamber pressure $(P_{amb})$ in which the given liquid is injected. $T_l$ is a superheated liquid temperature, and $T_{sat}$ is the saturated temperature. Blinkov et al. (1993) used the Blasius correlation to approximate the drag term in the correlation for bubble size ( $R_d$ ):

$$R_d = 0.58 K^{5/7} \left[ \left( \frac{\sigma R_{cr}}{\rho_l} \right)^{0.5} \left( \frac{\mu_l}{\tau_W} \right)^{7/10} \left( \frac{\rho_l}{\mu_l} \right)^{3/10} \right]^{5/7} \tag{2}$$

In eq. 2, $\rho_l$ is the liquid density, $\sigma$ is the surface tension, $\mu_l$ is the liquid viscosity and $\tau_w$ is the wall shear stress. The parameter $K$ in the equation considers the decrease in drag of a non-spherical bubble. The spherical shape of Lagrangian droplets restricts the parameter value to be unity in the present case. Since the first two wall nucleation models – the Riznic model and the Kolev model – are based on the sub-cooled wall boiling, they under-predict the nucleation rate for flashing flows. Further, the accuracy of the Blinkov model described above is proven for the flow through a convergent-divergent nozzle in the previous study (Janet et al. 2015). Hence, the Blinkov model is employed in the current study for simulating the wall nucleation at a low degree of superheat.

*1.1.2 Bubble Growth*

The bubble growth is best modeled by the general Rayleigh-Plesset equation (eq. 3) (Rayleigh 1917; Plesset 1949; Scriven 1960). This equation accounts for the inertial, pressure difference, and surface tension effects at various stages of bubble growth, as analyzed by Lee and Merte (Lee and Merte 1996) and in subsequent works (Lee and Merte 1996; Robinson and Judd 2004; Xi et al. 2017).

$$R\frac{\partial^2 R}{\partial t^2}+\frac{3}{2}\left(\frac{\partial R}{\partial t}\right)^2+\frac{2\sigma}{\rho_l R}+\frac{4\mu V_{bub}}{R}+\frac{4\kappa V_{bub}}{R^2}=\frac{P_v-P_l}{\rho_l} \tag{3}$$

Here, $R$ is the radius of the bubble at time instant $t$, $\mu$ and $\kappa$ are the surfaces normal and tangential viscosity respectively, $V_{bub}$ is the velocity of the bubble-liquid interface, $P_v$ and $P_l$ is the vapor and liquid pressures respectively. In the final stages of bubble growth, thermal effects start to dominate. Plesset and Zwick (1952) considered a boundary layer around the bubble and proposed a zero-order solution to the energy equation. Later, this model was modified to include vapor pressure effects (Plesset and Zwick 1954; Prosperetti and Plesset 1978). Mikic et al. (1970) solved energy and momentum energy with the Clausius-Clapeyron equation and the Rayleigh equation. Robinson and Judd (2004) analyzed all the terms and concluded that surface tension contributes significantly to the bubble growth at its inception. Then, the pressure difference accelerates the bubble-liquid interface to high velocities. At some point after growth, thermal and heat transfer effects start to take a lead role in bubble growth. In the initial stages of thermal effect-dominated growth, heat transfer takes place from the superheated bulk liquid across a thin thermal boundary layer into the bubble. As the bubble grows against the bulk liquid, thermal boundary layer thickness increases, in turn decreasing the heat transfer and subsequently slowing the bubble growth. For a detailed study on the cavitation and bubble dynamics, one can refer to the study of Christopher (1995). Similar to the methodology of Kato et al. (1994), Christopher (1995) used a critical time (as given by eq. 4), after which thermal effects are dominant than the inertial or pressure difference effects. The value of critical time decreases with an increase in temperature for a constant value of tension, indicating the initial set in of thermal effects.

$$t_c = \frac{(P_v - P_{amb})}{\rho_l} \cdot \frac{1}{\Sigma^2} \tag{4}$$

$$\Sigma(T_l) = \frac{{h_{fg}}^2 \rho_v^2}{\rho_l^2 c_{pl} T_l \alpha_l^{0.5}} \tag{5}$$

Here, $t_c$ represents the critical time during bubble growth, $P_v$ is the vapor pressure, $P_{amb}$ is the ambient pressure, and $\rho_l$ is the liquid density. The thermodynamic parameter $\Sigma(T_l)$ is the function of latent heat of vaporization $h_{fg}$, vapor-liquid density ratio ($\rho_v/\rho_l$), thermal diffusivity ($\alpha_l$), specific heat capacity ($c_{pl}$), and bulk temperature ($T_l$) of the liquid. Each model has its limitations and is valid at a particular stage of bubble growth. At initial stages, the general Rayleigh-Plesset equation is sufficient, while at final stages models considering thermal effects need to be used (Bhatia and De 2019). Christopher (1995)'s critical time definition (eq. 4 and eq. 5) is used for an appropriate choice of a model, which is discussed in detail in the results and discussion (section 5.1).

*1.1.3 Bubble burst*

It is the third and the last leg of flashing phenomena, which is also the least studied of all. The different aspects of the flashing-spray structure or any flashing system can be best approximated by modeling the bubble burst processes, such as velocities and size of the ejected droplets. Adachi et al. (1997) and Senda et al. (1994) modeled the number of child droplets being double the number of bubbles present inside a droplet. Zeng and Lee (2001) developed the break-up of superheated droplets based on the instability of the bubble-droplet system. All three models are based on the break-up of superheated droplets. Sher and Elata (1977) developed a model for the break-up of liquid core and estimated the ejected droplet size from equality of available work in the bubbles and the surface energy required to form the child droplets. Some of the disadvantages of this model lie in the fact that the bubble burst occurs at the moment when all bubble boundaries touch each other, which limits this model to predict any liquid core at a lower degree of superheat. Further, the model does not predict any ejection velocity and its direction inhibiting its use for the study of spray structure at higher superheat temperatures. Hence a lack of a complete physics-based model for a flashing spray is a motivation for the current work.

## 2. Mathematical Model for Bubble Burst

The bubbles are the vapor cavities with pressure and temperature corresponding to the superheated conditions of the surrounding liquid, and the bubble burst phenomena are common for bubbles to exist near the free surface. The cavity pressure is equal to the saturated pressure for the given superheated temperature of the liquid. The high vapor pressure inside the bubble pushes it against the surrounding gaseous medium at a lower ambient pressure. This sudden expansion of the bubble boundary moves the thin layer of liquid present between the free surface and the bubble. As evident in the experiments conducted by Blake and Gibson (1981) and later validated by Longuet-Higgins (1983), the bubble accelerates the thin layer of fluid to form a spherical dome. Along the center-axis of the dome emerges a jet, with the shape of approximately a circular cone. The similar phenomenon is observed in the case of the bubble burst at the ocean surface, resulting in the formation of sea-salt aerosol (Blanchard and Woodcock 1980).

Longuet-Higgins (1983) used the hyperbolic form of the Dirichilet's ellipsoidal flows (Lamb 1932) to model this liquid flow as the bubble grows near the free-surface. In this study, this model is extended for the flashing jets to predict the bubble burst process and the ejected droplets' properties. The concerned equations of the model are re-written for the convenience of understanding. The liquid flow is assumed to be inviscid, incompressible, and irrotational.

The velocity potential and the velocity vector in the Cartesian coordinate system is written as,

$$\phi = \frac{1}{2}\left( \frac{\dot{a}}{a}x^2 + \frac{\dot{b}}{b}y^2 + \frac{\dot{c}}{c}z^2 \right) \tag{6}$$

$$\nabla\phi = \left( \frac{\dot{a}}{a}x, \frac{\dot{b}}{b}y, \frac{\dot{c}}{c}z \right). \tag{7}$$

where $a, b, c$ are parameters, which are the functions of time only, and a dot ( $\dot{}$ ) represents time-differentiation.

The continuity equation with the velocity vector gives

$$\frac{\dot{a}}{a} + \frac{\dot{b}}{b} + \frac{\dot{c}}{c} = 0\,, \tag{8}$$

such that, $$abc = 2L^3 \text{ (constant)} \tag{9}$$

where $L$ represents the length parameter defined below.

At the scale of bubble size observed in flash-boiling sprays, gravity effects are negligible. For gravity to be significant in bubble evolution and bursting, the bubble size needs to be of the order of its characteristic scale (San Lee et al. 2011; Brasz et al. 2018), which is quantified by the non-dimensional number ( $Bo = \rho g R^2 / \sigma$ ) . The Bond number ( $Bo$ ) for the bubble size in flashing sprays is $\ll 1.0$ . Further, to simplify the equation, the surface tension effects are also neglected in the dynamics of a bubble bursting. For a homogeneous fluid of negligible gravity and unit density, the hydrodynamic pressure $P_{hyd}$ at any point is

$$-2P_{hyd} = 2\phi_t + (\nabla\phi)^2 + F(t) = \left( \frac{\ddot{a}}{a}x^2 + \frac{\ddot{b}}{b}y^2 + \frac{\ddot{c}}{c}z^2 \right) + F(t) . \tag{10}$$

Also $$-2\frac{DP_{hyd}}{Dt} \equiv \left( \frac{\partial}{\partial t} + \nabla\phi.\nabla \right)\left(-2P_{hyd}\right) = \frac{a\dddot{a} + \dot{a}\ddot{a}}{a^2}x^2 + \frac{b\dddot{b} + \dot{b}\ddot{b}}{b^2}y^2 + \frac{c\dddot{c} + \dot{c}\ddot{c}}{c^2}z^2 + \dot{F}(t) . \tag{11}$$

Here, $F(t)$ is a time varying function. The boundary conditions are taken such that $P_{hyd}$ and $DP_{hyd}/Dt$ vanish on the same free-surface (Longuet-Higgins 1972). Thus for the proportionality of the coefficients in equations (10) and (11),

$$\frac{\dddot{a}}{\ddot{a}} + \frac{\dot{a}}{a} = \frac{\dddot{b}}{\ddot{b}} + \frac{\dot{b}}{b} = \frac{\dddot{c}}{\ddot{c}} + \frac{\dot{c}}{c} = \frac{\dot{F}}{F} . \tag{12}$$

Integrating the equation (12) gives $a\ddot{a} = K_1 F,\ b\ddot{b} = K_2 F,\ c\ddot{c} = K_3 F$ , where $K_1,\ K_2,\ K_3$ are the constants. A change in scale of *a, b or c* does not alter the velocity potential, thus without the loss of generality we can take,

$$a\ddot{a} = -b\ddot{b} = -c\ddot{c} = -F . \tag{13}$$

From the above equation (10) and (13), the hyperbolic-shaped free surface ( $P = 0$ ) of the liquid core immediately above the bubble is given by,

$$\frac{x^2}{a^2} - \frac{y^2}{b^2} - \frac{z^2}{c^2} = 1, \tag{14}$$

which is a hyperboloid with semi-axes $a, b, c$.

In case of an axisymmetric hyperboloid, the equation of free-surface hyperboloid becomes,

$$\boldsymbol{\frac{x^2}{a^2} - \frac{y^2 + z^2}{b^2} = 1} \tag{15}$$

and similarly, the equation of the bubble hyperboloid (inner hyperboloid) belonging to the same family of hyperboloid curves is,

$$\frac{x^2}{a^2}-\frac{y^2+z^2}{b^2}=A^2 \tag{16}$$

To derive the velocity of the axisymmetric hyperboloid vertex, continuity equation can be re-written as

$$\dot{a}\ddot{a}-2\dot{b}\ddot{b}=0,$$

the integration of which leads to a constant,

$$\dot{a}^2-2\dot{b}^2=U^2. \tag{17}$$

With the dimensionless parameters $\alpha=a/L$, $\beta=b/L$ and $\tau=Ut/L$, equation (9) becomes

$$ab^2=2L^3. \tag{18}$$

In non-dimensional form, $\alpha\beta^2=2$. (19)

The velocity parameter U is directly proportional to bubble velocity $U=C_1V_{bub}$, while the length parameter is proportional to the depth of the bubble from the free surface $L=C_2R_{M,S}$, where $R_{M,S}$ is the depth of the bubble from the free surface. The constants are fixed based on the agreement with the experimental results (Kamoun et al. 2010; Lamanna et al. 2014), such that $C_1=12400/P_{amb}^{0.85}$ and $C_2=0.02$. From equation (17) and (18), the velocity of a particle at a vertex $=(\dot{a},0,0)$ is obtained as,

$$\dot{a}=\frac{U}{\left(1-\alpha^{-3}\right)^{\frac{1}{2}}}=V_{eject}. \tag{20}$$

It provides the maximum velocity possibly achievable by a particle. Since the ejected droplets with high transverse velocity form the periphery of the spray-structure, it is assumed that the ejected droplet has velocity equal to the vertex of the free surface hyperbola.

From equation (13) $$F=-a\ddot{a}=-U^2\alpha\frac{\partial\alpha}{\partial\tau^2}=\frac{\frac{3}{2}U^2\alpha^{-3}}{\left(1-\alpha^{-3}\right)^2}. \tag{21}$$

The above equations define the structure of the hyperboloids essentially based on non-dimensional parameters $\alpha$ and $\beta$ at a particular time instant $\tau$ . The evolution of $\alpha$ (when $\alpha > 1$) with $\tau$ is specified by,

$$\tau = \int_1^\alpha \left(1 - \frac{1}{2}\alpha^{-3} - \frac{1}{8}\alpha^{-6} - \ldots\right) d\alpha \,. \tag{22}$$

$$\text{or approximately } \alpha = (\tau - \tau_0) - \frac{1}{4}(\tau - \tau_0)^{-2} \,. \tag{23}$$

$$\text{When } \alpha \to 1, \qquad \tau = \int_0^\eta \left[1 - (1+\eta)^{-3}\right]^{\frac{1}{2}} d\eta \sim \frac{2}{\sqrt{3}}\eta^{\frac{3}{2}} \tag{24}$$

For this range of $\alpha$ , the dependency on $\tau$ is given by $\alpha \sim 1 + \left(\frac{\sqrt{3}}{2}\tau\right)^{\frac{2}{3}}$ . (25)

The growth of the Dirichlet hyperboloids starts when the limiting angle between the asymptotes is $\approx 109.47^o$ as found by Longuet-Higgins (1983). It is assumed that once the bubble size increases to reach the free surface, the asymptotes form the given limiting angle ($\approx 109.47^o$), initiating the formation of bubble hyperboloid. Before this instant and after the inception, the bubbles are assumed to be spherical in shape. Whereas, some portion of the bubble remains hyperboloid in shape after the instant of formation of critical angle, as described by equation (16).

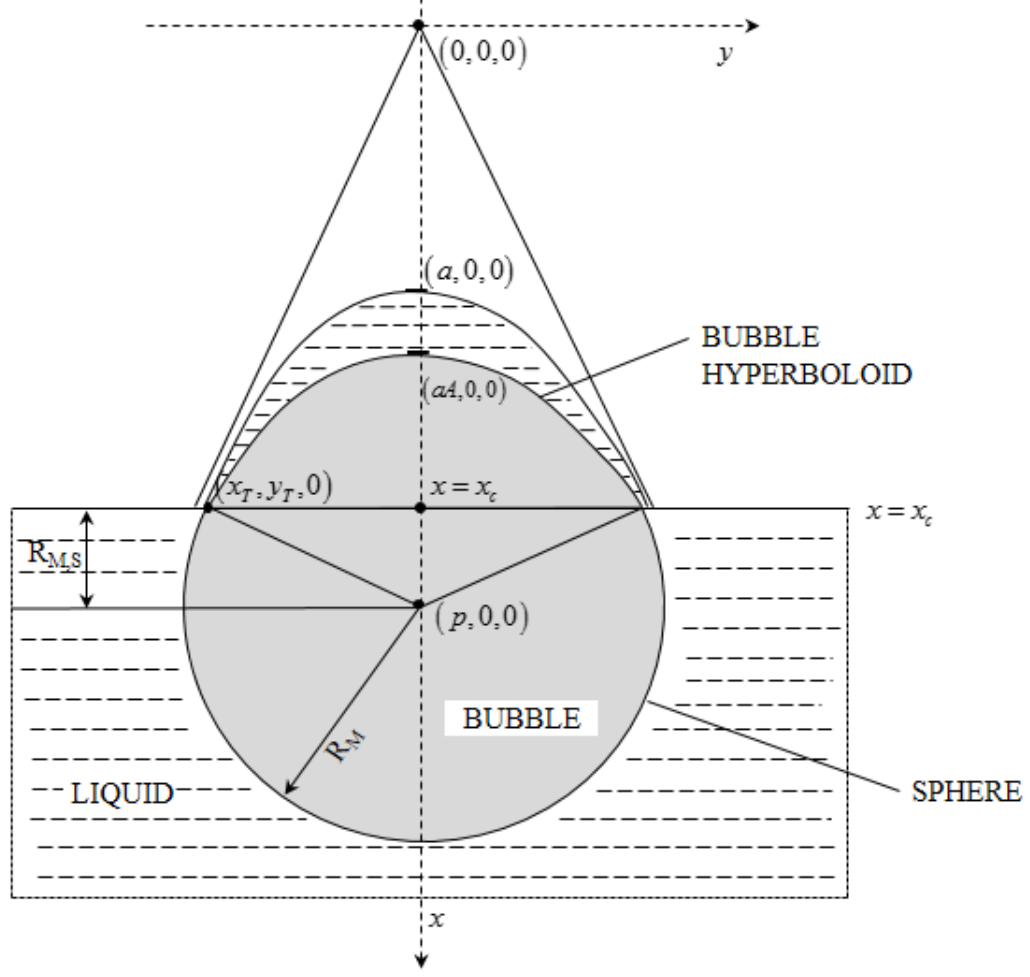

**Fig. 1.** Schematic of a superheated liquid-bubble system.

As illustrated in Fig. 1, a specific part of the bubble is hyperboloid in shape while the rest is spherical. Since fluid flow above the free surface of liquid jet-core is assumed to be hyperbolic, a relation between the time-varying hyperboloid section and the spherical section of the bubble is developed to maintain the consistency. To solve the system, the center of hyperbolas is taken as a coordinate origin. From Fig. 1, the relation between the center of the bubble and the free surface level of the liquid can be described as,

$$x_c = p - R_{M,S} \tag{26}$$

Here, the distance of the free surface from the origin is $x_C$, the distance of the bubble center from the origin is $p$, and the depth of the bubble is $R_{M,S}$. Assuming the droplets are far from each other and do not interact in any way throughout the bubble evolution time, the value of $R_{M,S}$ is assumed constant ($\sim 1.6 \times 10^{-5}$) based on the experimental validations (Kamoun et al. 2010). Since the problem is axisymmetric about the x-axis, it can be solved in 2-dimensions, and the bubble can be represented as a circle. The hyperboloid remains tangent to the bubble at all times. In 2-dimensional terms, the hyperbola is tangent to the circle at a point $\left(x_T, y_T\right)$. Equating the slopes of the hyperbola and the circle at this point $(y'_{circle} = y'_{hyp})$, we get

$$x_T = \frac{p}{\left(\dfrac{\beta^2}{\alpha^2}+1\right)} \quad , \quad y_T = \sqrt{\frac{x_T^2}{\left(\dfrac{\alpha^2}{\beta^2}\right)} - b^2 A^2} \tag{27}$$

The center of the bubble $(p, 0,0)$ from the origin is located by substituting $x_T$ and $y_T$ in the two-dimensional bubble's equation,

$$p = \sqrt{\left(1+\frac{\alpha^2}{\beta^2}\right)\left(R_M^2 + b^2 A^2\right)} \tag{28}$$

*2.1. Break-Up Criteria*

It is an important condition that decides when the evolving layer of liquid (between the two surfaces of free-surface hyperboloid and bubble hyperboloid) will break away from the parent parcel and form a child droplet. At the scale of the bubble-liquid system, surface tension plays a significant role along with the inertial and pressure difference terms. This is evident from the high values of Laplace number ( $\rho\sigma L / \mu^2 > 1.0$ ) for the scale of bubbles available in the flashing sprays. On the one hand, the energy due to inertia and pressure difference enhances the break-up, while the surface tension, on the other hand, opposes it.

The break-up process will proceed if the following condition is satisfied,

$$e_s \le e_p + e_k \tag{29}$$

Here $e_s$ is the surface energy, $e_p$ which is the pressure potential energy due to pressure difference across the thin layer of liquid and $e_k$ is the kinetic energy of the liquid layer at any moment before the bubble burst. For calculation of surface energies, the surface area of the hyperboloids defined by equation (15), (16), and the spherical section of the bubble is calculated.

$$SA_1 = \frac{2\pi b}{a^2}\sqrt{a^2+b^2}\left[\frac{x_c}{2}\sqrt{x^2-c^2} - \frac{C^2}{2}\ln\left(x_c+\sqrt{x_c^2-c^2}\right) - \frac{a}{2}\sqrt{a^2-c^2} + \frac{C^2}{2}\ln\left(a+\sqrt{a^2-c^2}\right)\right], \tag{30}$$

$$SA_2 = \frac{2\pi b}{a^2}\sqrt{a^2+b^2}\left[\frac{x_c}{2}\sqrt{x_c^2-c_2^2}\right.$$

$$\left. - \frac{C_2^2}{2}\ln\left(x_c+\sqrt{x_c^2-c_2^2}\right) - \frac{aA}{2}\sqrt{a^2A^2-C_2^2} + \frac{C_2^2}{2}\ln\left(aA+\sqrt{a^2A^2-C_2^2}\right)\right] \tag{31}$$

$$SA_3 = 2\pi\left[R^2{}_M + \frac{R_M b^2 p}{b^2+a^2}\right] \tag{32}$$

Here, $SA_1$ and $SA_2$ are the surface area of hyperboloids. $SA_3$ is the surface area of the spherical section of the bubble system, as shown in Fig. 2. Also, the constants are $C_1 = a^4/a^2+b^2$, and $C_2 = a^4A^2/a^2+b^2$. The energy variation leading to the bubble burst is discussed in detail in Results section 5.

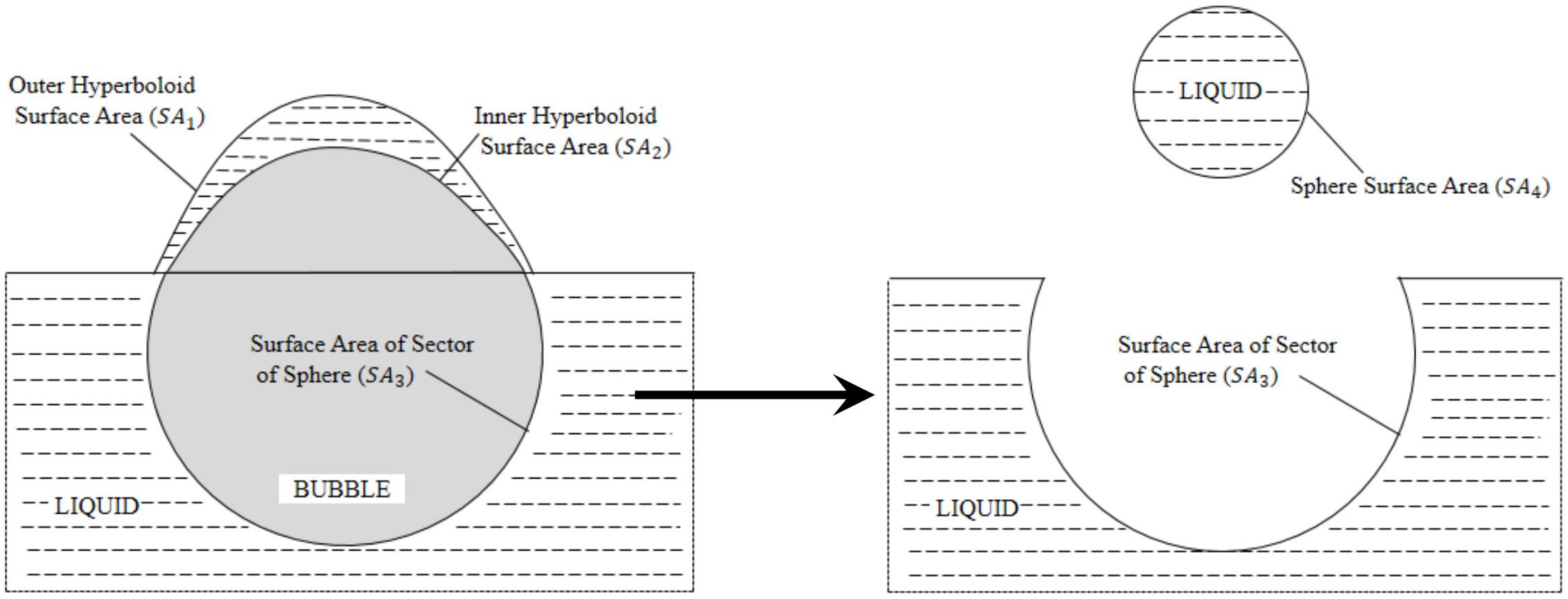

**Fig. 2.** Schematic of child droplet break-up process.

The mass of the ejected droplet is equal to the product of the liquid density and the volume of the layer of liquid between two hyperboloids, which is given by,

$$m_{eject} = \frac{\rho_l \pi B^2 L^3}{3}\left[3A^2\left(\frac{x_c}{L}\right) - 2A^3\alpha - 3\left(\frac{x_c}{L} + 2\alpha\right)\right] \tag{33}$$

## 3. Test Case Description

The current flash-boiling model is validated for three different fuels at different pressure and superheats. The details of the test cases are based on the experimental cases (Kamoun et al. 2010) and given in table 1.

| | Fuels | | |
|---|---|---|---|
| | Ethanol | Acetone | Iso-Octane |
| $P_{amb}(kPa)$ | $\Delta T(K)$ | $\Delta T(K)$ | $\Delta T(K)$ |
| 6000 | 12, 20, 25, 34 | - | - |
| 10000 | 12, 20, 25, 34 | 12, 20, 27, 34, 44, 54 | 12, 15, 25, 34, 40, 50, 65 |
| 20000 | 12, 20, 25, 34, 42 | 12, 20, 27, 34, 44, 54 | 12, 15, 25, 34, 40, 50 |

**Table 1.** Test cases of three different fuels at various chamber pressures ( $P_{amb}$ ) and degree of superheats ( $\Delta T$ ).

In work by Girsheck and Chiu (1990) and later by Kamoun et al. (2010) and Lamanna et al. (2014), a non-dimensional form of surface tension is a focus which is also a part of their new “homogeneous” nucleation rate formula. It is stated in Kamoun et al. (2010) that the surface tension of the superheated liquid, and not the degree of superheat decides the transition condition for the onset of fully flashing regime. Since the present work concentrates on the transition regime of flashing sprays with the degree of superheat less than the spinodal limit, a scant amount of experimental data is available for validation of spray structure. The data accessible in the public domain is that of the Kamoun et al. (2010). Though this set of experimental data is mostly in the fully flashing regime, part of it is available in the transition regime for ethanol sprays. The transition regime is decided upon by the empirical correlations of Cleary et al. (2007).

$$\begin{aligned} &\text{Start of Transition Regime: } Ja\Phi = 55 We_v^{-1/7} \\ &\text{End of Transition Regime: } Ja\Phi = 150 We_v^{-1/7} \end{aligned} \tag{34}$$

Here, the non-dimensional degree of superheat is represented by a non-dimensional Jacob number $(Ja = \rho_l c_{pl} \Delta T / \rho_v h_{fg})$ , whereas the non-dimensional velocity is denoted by a non-dimensional Weber number $(We_v = \rho_v \mathbf{u_0^2} D / \sigma_l)$ (Cleary et al. 2007). Here, $D$ represents the nozzle diameter, $\sigma_l$ is the liquid surface tension, $\mathbf{u_0}$ is the inlet velocity of the injected liquid and $\Phi = 1 - \exp\left[-2300\left(\rho_v / \rho_l\right)\right]$ . In the current work, the superheat temperatures for the simulations are chosen based on this condition (Fig. 3(a)). Here, the *dotted line* is the starting point of the transition regime, whereas the *dashed* line is a point of onset of a fully flashing regime and end of the transition regime.

Instead of dependence on the dominance of flash-boiling break-up over a mechanical break-up, it is the kind of nucleation that decides the onset of a fully flashing regime[9]. It is this transition regime, where heterogeneous nucleation is dominant. Also, the spread-angle of the flashing spray continuously increases till the point in the degree of superheat (0.8-0.9 times the critical temperature[27]) when homogenous nucleation starts to flash the incoming jet more violently. Once crossing over to this fully flashing regime, the bell-shaped spray structure is formed, maintaining it throughout the rest of the superheat temperature range. The present case, as per Lamanna et al.'s (2014) onset criterion, is reported in Fig. 3(b). The data points chosen for the present study lies in or near the onset region of fully-flashing sprays. The criterion for the fully-flashing onset is given by a parameter $\chi$ dependent on non-dimensional surface tension $\Theta$ and pressure ratio $R_p$ :

$$\chi = \frac{4}{27} \cdot \frac{\Theta^3}{\left(\ln R_p\right)^2} . \tag{35}$$

Where $\Theta = a_0 \sigma / k_b T_{inj}$ , as given by Girshick and Chiu (1990) and $R_p = P_{sat} / P_{amb}$ . The parameter $k_b$ is the Boltzmann constant, $T_{inj}$ is the injection temperature and $a_0$ is the molecular surface area.

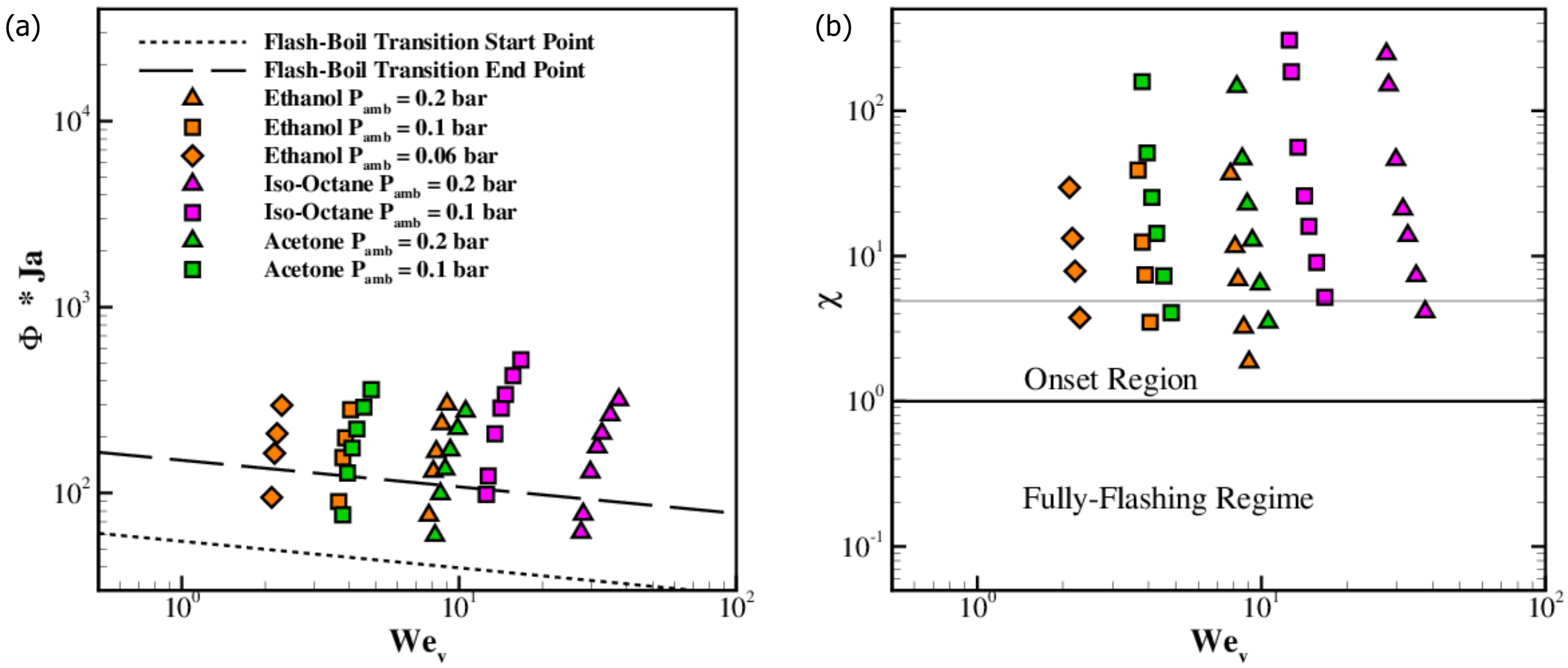

**Fig. 3.** Investigated cases in (a) the transition regime as defined by Cleary et al. (2007), (b) the onset of fully flashing regime as defined by Lamanna et al. (2014).

## 4. Numerical Implementation

The above-described model is implemented in the OpenFOAM framework (Greenshields 2015). It is in the Lagrangian framework where a complete three-dimensional simulation of a liquid spray can be carried out in a large domain. This allows the study of various aspects of spray structure, like spread angle, at any downstream distance from the injection nozzle.

### *4.1. Governing Equations*

The dispersed (liquid) phase is solved with the Lagrangian approach while the continuous phase is solved in the Eulerian framework. The governing equations solved for the gas phase are:

Continuity equation,

$$\frac{\partial \rho}{\partial t} + \nabla(\rho u) = S_{mass} \tag{36}$$

Momentum equation,

$$\frac{\partial \rho u_i}{\partial t} + \frac{\partial}{\partial x_j}(\rho u_i u_j) = -\frac{\partial p}{\partial x_i} + \rho g_i + \frac{\partial}{\partial x_j}\left(\tau_{ij}\right) + S_{mom,i} \tag{37}$$

Species equation,

$$\frac{\partial \rho Y_k}{\partial t} + \frac{\partial}{\partial x_j}(\rho Y_k u_j) = \frac{\partial}{\partial x_j}\left(\rho D_k \frac{\partial Y_k}{\partial x_j}\right) + S_{species} \tag{38}$$

Energy equation,

$$\frac{\partial \rho h}{\partial t} + \frac{\partial\left(\rho h u_j\right)}{\partial x_j} = \frac{DP}{Dt} + \frac{\partial}{\partial x_j}\left(q_j\right) + \tau_{ij}\frac{\partial u_i}{\partial x_j} + S_{enthalpy} \tag{39}$$

Here, $\rho$ is the gas-phase density, $u_i$ is the gas-phase velocity along $i^{th}$ direction, $Y_k$ is the mass-fraction of $k^{th}$ species and $h$ is the enthalpy. The viscous heating term being smaller than other terms in the above equation, is neglected. The viscous stress tensor $\tau_{ij}$ is defined as

$$\tau_{ij} = 2\mu\left(S_{ij} - \frac{1}{3}\delta_{ij}S_{ii}\right) \tag{40}$$

where, $S_{ij} = \frac{1}{2}\left(\frac{\partial u_i}{\partial x_j} + \frac{\partial u_j}{\partial x_i}\right)$. (41)

The Lagrangian particles, which serve as liquid droplets, are treated as a point source of mass, momentum, and energy for the above equations. Various models for different phenomena like dispersion, collision, atomization, heat transfer, etc., are solved for each particle to replicate the real case. The Basset-Boussinesq-Ossen (BBO) equation is solved for momentum conservation of the Lagrangian particles. The equation is given by,

$$\frac{d\vec{X}_p}{dt} = \vec{U}_p \tag{42}$$

$$\frac{d\vec{U}_p}{dt} = \frac{\vec{U}_{g,rel} - \vec{U}_p}{\tau_p} + \vec{g} \tag{43}$$

Here $\tau_p = \tau_p^{St} / C_D$ is the droplet relaxation time for Stokes flow. The particle displacement and the velocity is represented by $X_p$ and $U_p$ whereas $U_{g,rel}$ is the relative velocity with respect to the gaseous phase, $g$ is the gravitational acceleration. The drag on the particles is approximated by the Schiller-Naumann correlation for the Stokes drag:

$$C_D = \begin{cases} \frac{24}{\mathrm{Re}_p}\left(1 + 0.15\,\mathrm{Re}_p^{0.687}\right) \\ 0.44 \end{cases}, \tag{44}$$

where $\mathrm{Re}_p$ is the particle Reynolds number. The Kelvin-Helmholtz and Rayleigh-Taylor models are used to describe the aerodynamic break-up. The NSRDS – AICHE database of National Institute Standards and Technology (Daubert 1989) is used to calculate the thermodynamic properties for the two phases at different conditions of pressures and temperatures.

*4.2. Post-Processing*

Apart from the visual qualitative assessment of flashing sprays, the spread-angle of a spray is an acceptable measure of spray structure. The sprays being prone to asymmetricity along the axis, instead of a more intuitive $\theta_z = 2\tan^{-1}\left(r_{\max}/z\right)$ formula, the spread angle at any position ‘z’ mm downstream of the injector is measured by the following method (Fig. 4):

$$\theta_z = \tan^{-1}\left(\frac{x_{\max}}{z}\right) + \tan^{-1}\left(\frac{|x_{\min}|}{z}\right). \tag{45}$$

Here, $x_{\max}$ is the x-coordinate of the droplet farthest away from the jet centerline along the positive x-axis, and $|x_{\min}|$ is the x-coordinate of the droplet farthest away from the centerline along negative x-axis.

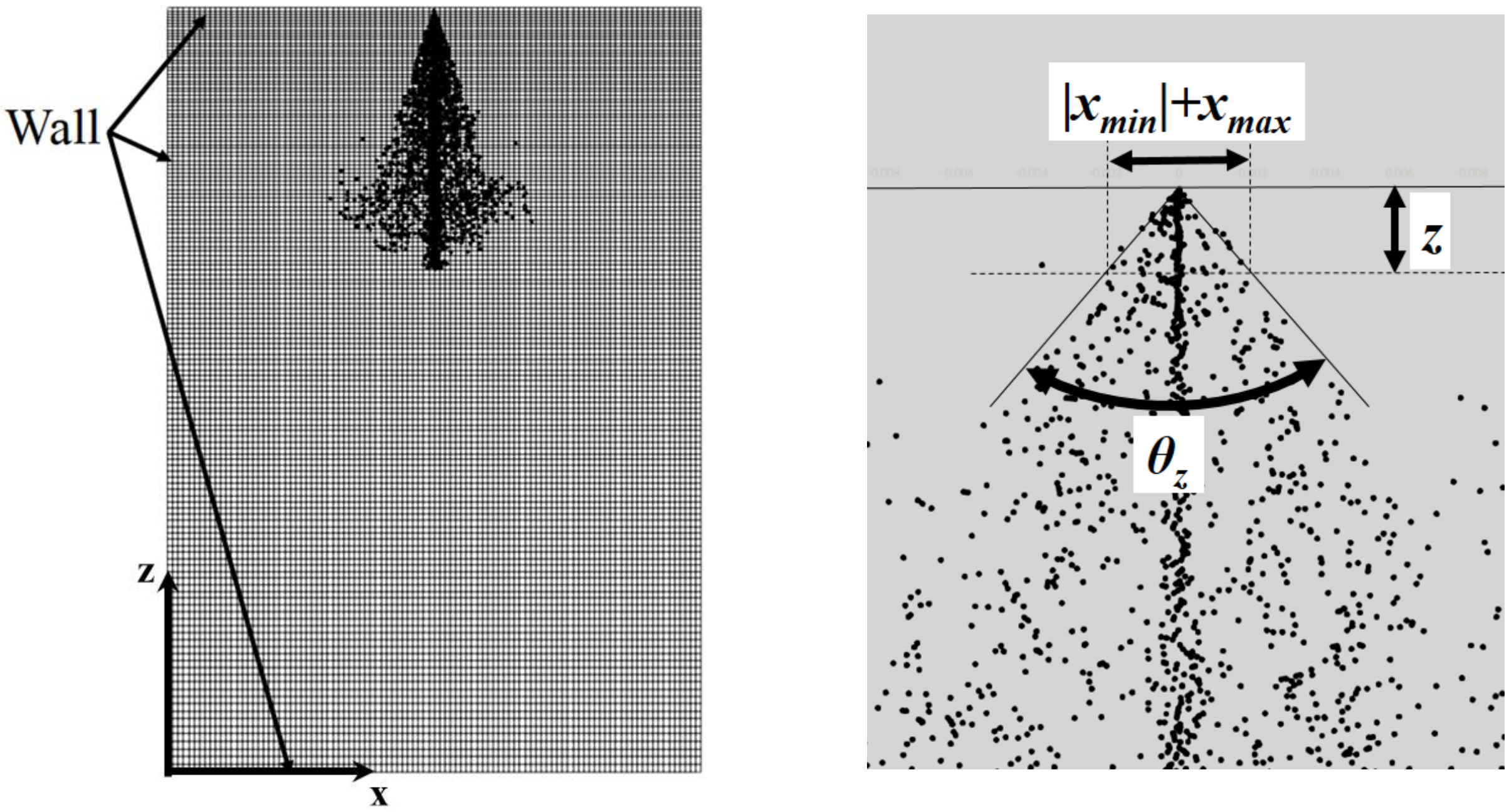

(a) (b)

**Fig. 4.** (a) Schematics of Computational Domain, (b) Spread-angle measured for sprays approximated by the Lagrangian droplets.

*4.3 Computational Details*

The computational domain is a chamber 150 mm high and 100 x 100 mm$^2$ cross-section with the wall as the boundary condition. The liquid is injected from the top of the computational domain through a nozzle of diameter ($d_n$) 150μm, as illustrated in Fig. 4(a). The bottom wall being $1000 \cdot d_n$ downstream of the nozzle while the side walls $333 \cdot d_n$ far from the nozzle location have an insignificant effect on the near-nozzle spray region. As discussed by Park and Lee (1994), the spray region within 40 nozzle diameters downstream is unaffected by the entrainment air. Hence, the chamber geometry of the above size is going to have very little effect, if any, on the near-nozzle spray structure. The model calculations, as mentioned above, are independent of the grid cell or the domain size. This results in the ejected droplet size and the velocities also being independent of the grid size. Yet, the validation for the consistency of spray structure is carried out on the grids of three different sizes – fine (150x150x200), medium (100x100x150), and fine (80x80x120). The grid independence test is carried at four different degrees of superheat (12K, 20K, 25K, and 34K) for ethanol fuel. As shown in Fig. 5, the spray angles measured for the three grid sizes are not only consistent with each other but also with the fact that the model is independent of the grid. Hereon, the simulations are based on the medium grid, the results of which are validated and analyzed in the following sections.

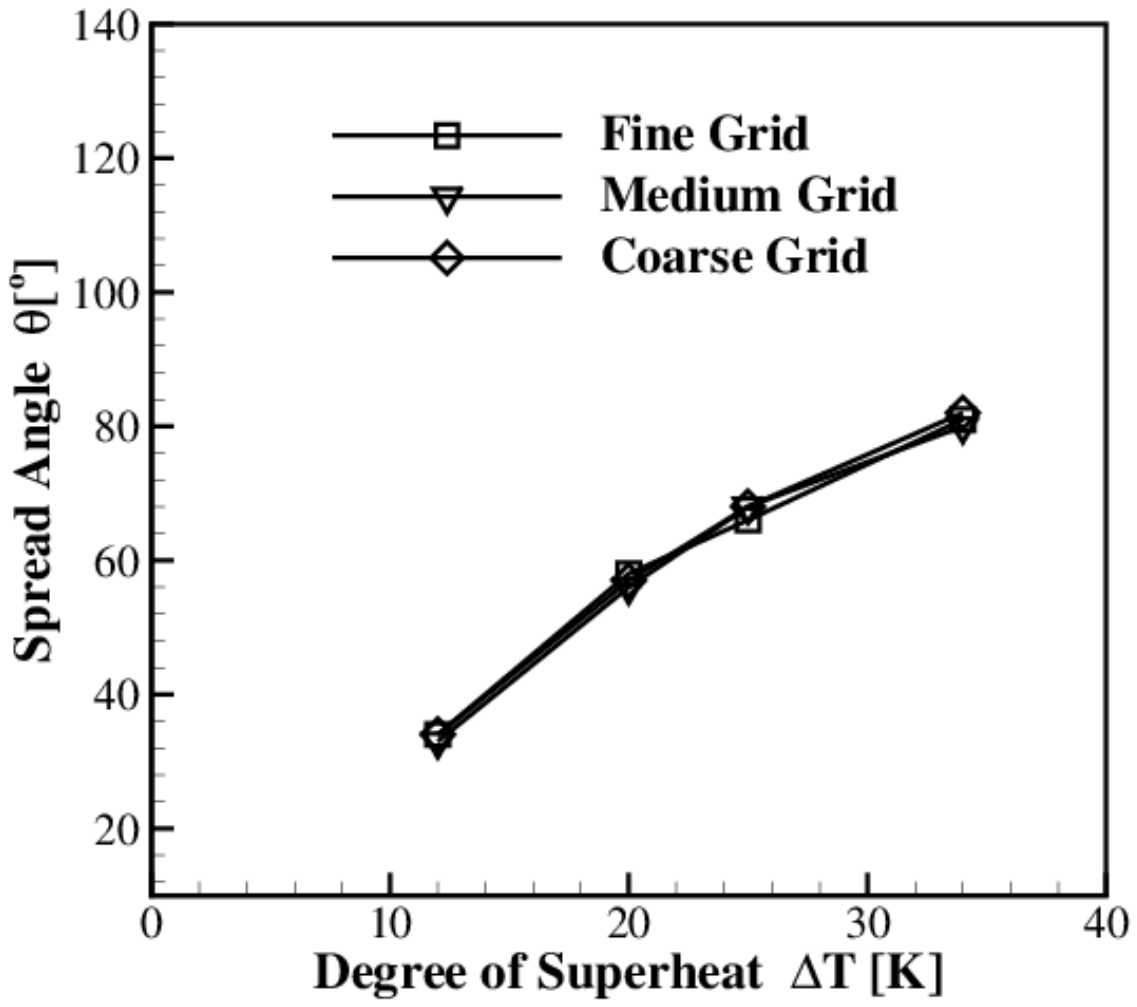

**Fig. 5.** Spread angle versus superheat degree for three grid sizes – fine (150x150x200), medium (100x100x150) and fine (80x80x120), for ethanol fuel at 10kPa ambient pressure

## 5. Results and Discussion

The open-source code is validated in the previous works (Ghasemi Khourinia 2017; Bhatia and De 2018) for its different modules of Lagrangian-Eulerian based solvers. The previous validations by the authors included that of the liquid penetration depth, Sauter mean diameter (SMD) of droplets and qualitative spray structure.

The simulations are carried for the test cases given in table 1, and the spray spread-angles measured, is validated against the detailed data of the spread-angle presented by Kamoun et al. (2010) at a lower superheat of ethanol.

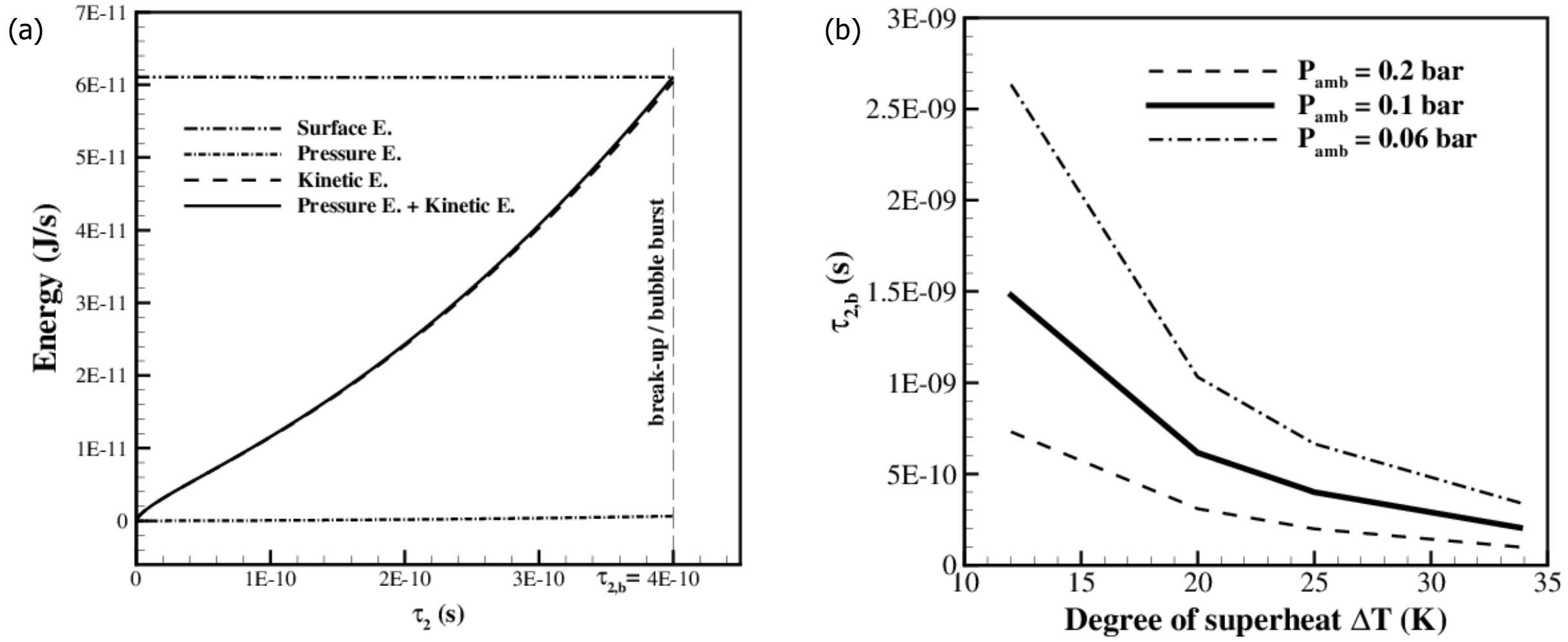

**Fig. 6.** (a) The evolution of pressure, surface and kinetic energy with time $\tau_2$ , (b) Break-up time $\tau_{2,b}$ at different superheat degrees and ambient pressures

*5.1. Bubble-burst time*

Before moving ahead with the discussion of the spread-angle, it is necessary to analyze the break-up criteria term-by-term in the given conditions. This gives the physical understanding as of what is happening when bubble size increases and bulges the liquid layer (between the free-surface and bubble hyperboloids) just above it and consequently ejecting it at a particular moment. Equation 29 compares the magnitude of driving forces – pressure energy, surface energy, and kinetic energy – and there relative dominance in the whole process of the bubble burst. It is worth mentioning that the bulge of the liquid layer starts to form only once the bubble reaches the free-surface. Thus, if $\tau_1$ is the time taken for the bubble to grow till it reaches the free-surface and $\tau_{2,b}$ time to form a bulge of liquid layer till bubble burst, then the total time to bubble-burst from the inception of the bubble is

$$\tau_{bub} = \tau_1 + \tau_{2,b} \tag{46}$$

Fig.6(a) reports the time evolution of the three dominant energies after the bulge is formed until the bubble burst for ethanol fluid at a chamber pressure of 10kPa and 25K superheat degree. As mentioned earlier, potential energy due to pressure difference across the liquid layer and the kinetic energy of the layer together assist in its break-up from the liquid-core. Hence, when the sum of pressure and kinetic energy becomes more dominant than the energy due to surface tension, the bubble burst occurs (eq. 29). According to Fig. 6(a), the bubble burst happens at $4\text{x}10^{-10}$ s after its formation of liquid layer bulge. As observed, the minuscule change in pressure potential energy while the exponential increase in kinetic energy with time can be explained by the fact that the potential energy due to pressure difference across the liquid layer continuously changes into the kinetic energy of the liquid layer. The kinetic energy almost overlaps with the total assisting force (pressure energy + kinetic energy). It can be said that kinetic energy and energy due to surface tension effects play a significant role in determining the time of burst.

| **Degree of Superheat (K)** | **Critical Time $t_c$ (μs)** | **Break-up time $\tau_{bub}$ (μs)** |
|---|---|---|
| 12 | 350 | 9.3 |
| 20 | 750 | 5.5 |
| 25 | 1100 | 4.3 |
| 34 | 2000 | 2.9 |

**Table 2.** Break-up time and Critical time for ethanol flashing jets at ambient pressure of 10kPa.

When the bubble-burst time is compared for different superheat degrees and pressure conditions in Fig. 6(b), it is found that $\tau_{2,b}$ decreases with an increase in superheat degree at a constant pressure. This can be attributed to the higher velocity of the liquid layer (between free-surface and the bubble), which is the result of high bubble-liquid interface velocity when the bubble touches the free surface. Thus, the break-up condition reaches earlier for higher superheat degree cases.

The total break-up time $\tau_{bub}$ also follows the same decreasing trend with the increase in the degree of superheat as $\tau_{2,b}$ shown in table 2. Further, the comparison of $\tau_{bub}$ with Christopher (1995)'s critical time (as described by eq. 4 in section 1.1.2) in table 2 gives a fair idea of the driving force of the bubble growth. The $\tau_{bub}$ being smaller than $t_c$ by hundreds of order of magnitude shows that the inertial and pressure difference controls the bubble growth. Hence, the modeling of the thermal effects can be safely ignored for the bubble growth stage. Also, $t_c$ is found to increase with the degree of superheat, contrary to the trend of $\tau_{bub}$, due to the fact that $t_c$ is directly proportional to the pressure difference across the bubble-liquid interface as given in eq. 4. Thus, as the degree of superheat is increased, the vapor pressure also increases, and with no change in the ambient pressure, an increase in $t_c$ is subsequently witnessed.

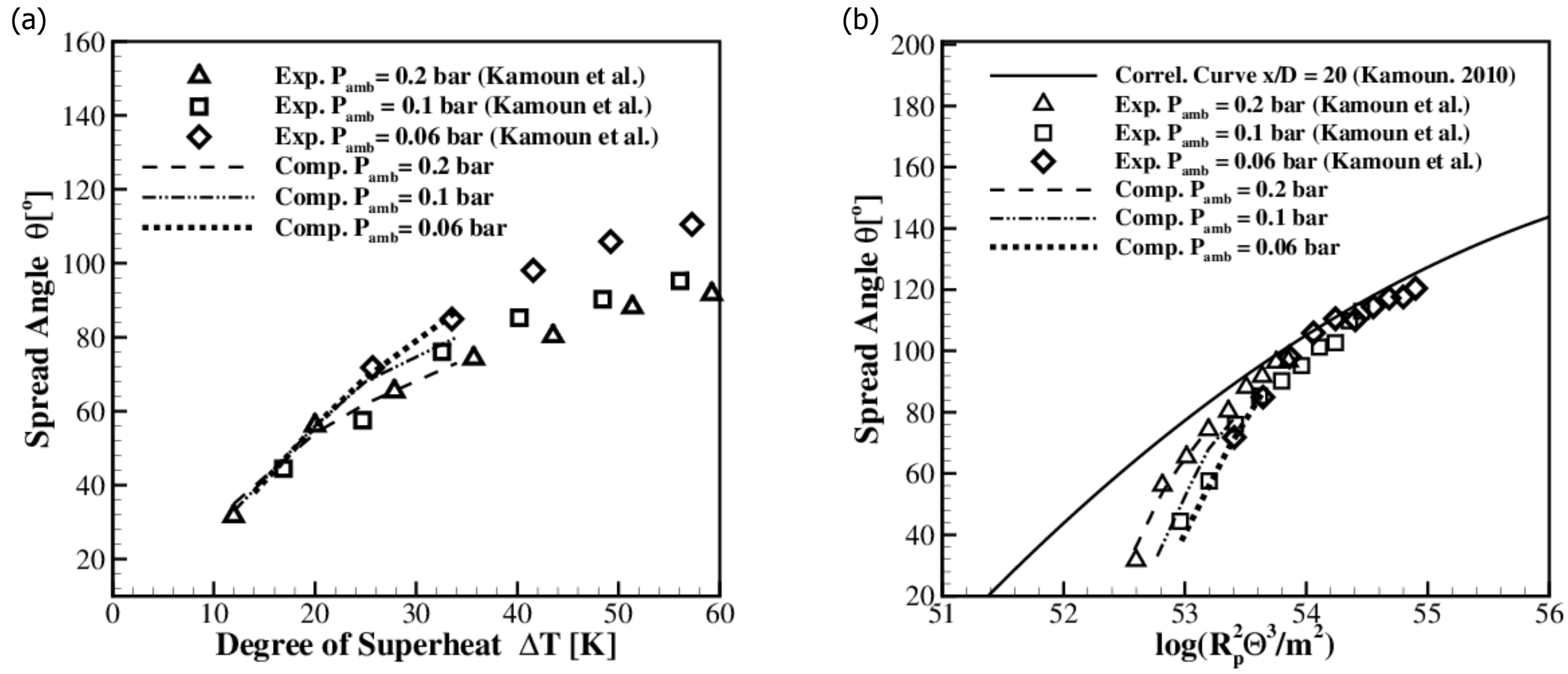

**Fig. 7.** (a) Spray spread-angle at different superheat degrees and ambient pressures, (b) spray spread-angle with respect to $\xi$ compared to the experimental results of Kamoun et al. (2010) at x/D = 20.

*5.2. Spread-Angle*

The results of the spread-angle, as predicted by the model at a downstream distance of x/D = 20 for various chamber pressures for ethanol, are depicted in Fig 7(a). The correlations given by Lamanna et al. (2014) at different downstream distances provide a universal tool to predict the spread angle in the fully-flashing regime for various fluids. As identified in their study and discussed above, it is the type and intensity of nucleation that decides the shape of spray structure, not the superheat degree.

Based on this, Lamanna et al. (2014) used an expression ( $\xi = \log(R_p^2\Theta^3 / m^2)$ ) of non-dimensional parameters to develop the spread angle correlation. Only valid in a fully flashing regime, the correlation at a lower superheat degree exhibit considerable deviation from the experimental results (Fig.7(b)).

The spread angle versus $\xi$ is steeper at a lower superheat where the presence of liquid jet-core allows the use of the proposed model. The current data is, thus, compared against the experimental data (Kamoun et al. 2010) and the correlation curve (Lamanna et al. 2014) in Fig. 7(b). As visible in Fig. 7(b), the experimental data for different chamber pressures overlaps and falls on the same curve. Well, this indicates that $\xi$ is indeed a parameter that controls spray structure at a specific downstream distance from the nozzle. At lower superheats in the range where liquid jet-core exists, the behavior of spray changes and the results are well predicted by the proposed model. Here, the liquid jet-core allows the bubble to traverse its length and grow to a point where the bubble-burst happens, and the child droplets are ejected radially. These large radial-velocity droplets are soon slowed down due to surrounding gas resistance. This gas resistance is critical in transforming a perfectly conical-shaped spray to a bell-shaped spray structure at the end of the transition regime or the start of the fully-flashing regime, especially for the externally flashing sprays. Fig. 8(a) and Fig. 8(b) verify it, as an almost perfect conical spray at the degree of superheat of 12K, and pressure 10kPa becomes into a bell-shaped structure at 34K of superheat. Lower the chamber pressure (i.e., higher the Rp), less is the surrounding gas resistance, farther it would travel away from jet center-axis, and higher would be the spray spread-angle. This observation is apparent in Fig. 7(a) where at the same degree of superheat, the spray-spread angle increases with lowering the chamber pressure.

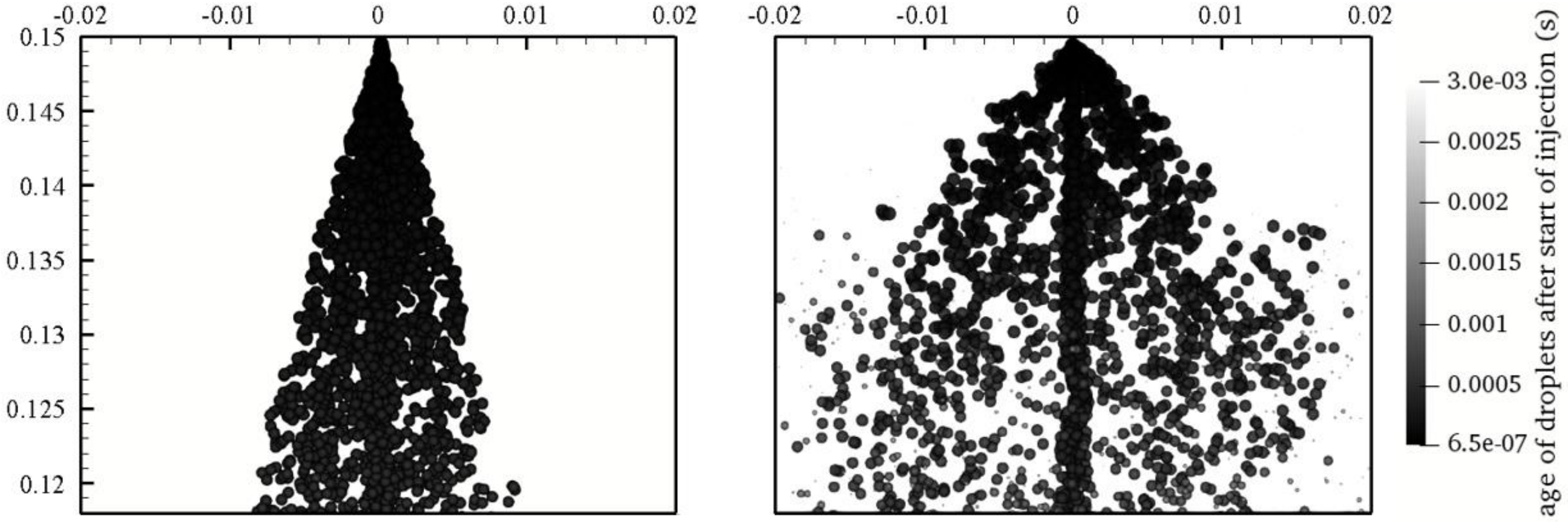

**Fig. 8.** Lagrangian spray at the degree of superheat (a) 12K and (b) 34K with the ambient pressure of 10kPa.

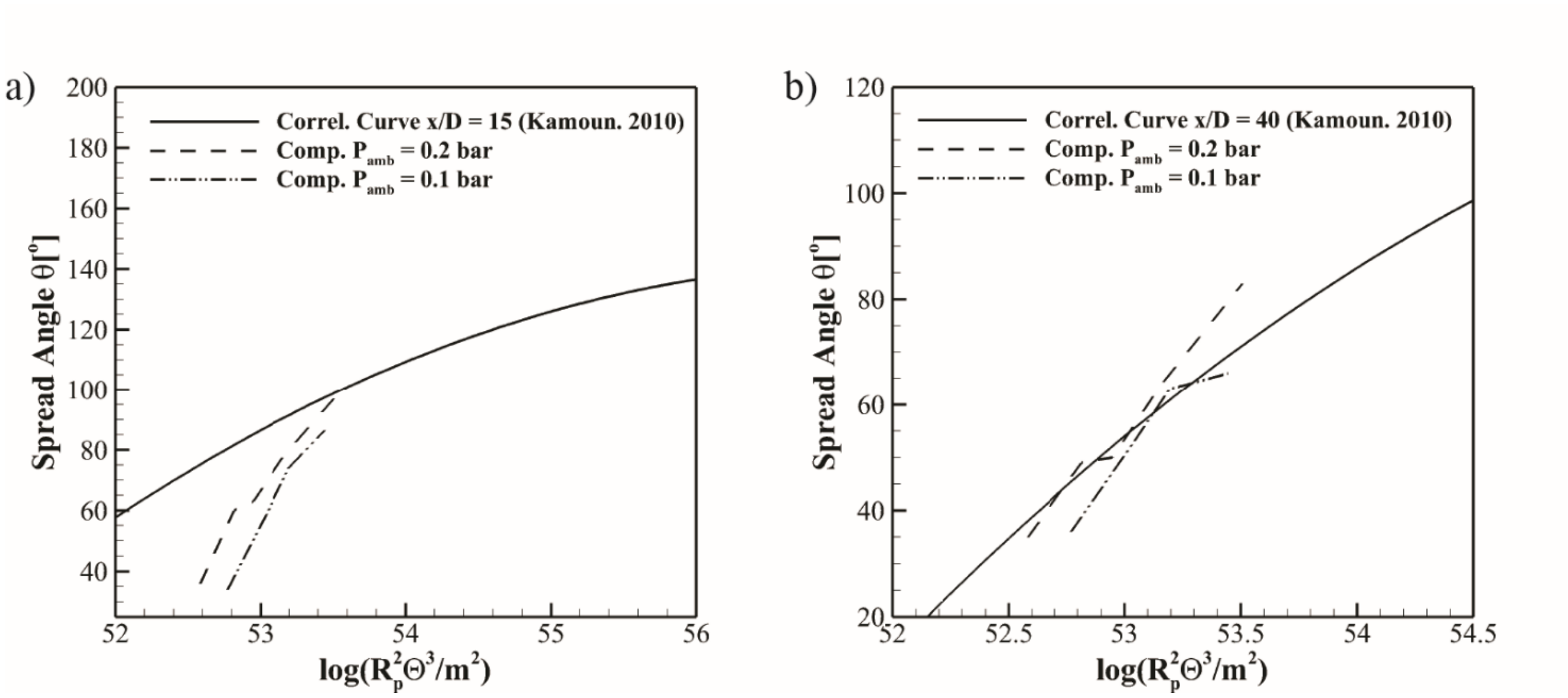

**Fig. 9.** Spray spread-angle of ethanol with respect to $\xi$ at (a) x/D = 15 and (b) x/D = 40

The spread-angle, as predicted by the current model, is also reported against the experimental data and correlation curve (by Kamoun et al. (2010)) at x/D=15 and x/D=40 in Fig. 9(a) and Fig. 9(b), respectively. As noted in the works of Park and Lee (1994), Lamanna et al. (2014) and Nagai et al. (1985), the spread angle beyond 40 nozzle diameters downstream (x/D>40) is affected by the air entrainment effects and geometry of the confined chambers. This is evident from Fig. 8(b) where the spray starts to collapse in the downstream region. Since the proposed model is valid for the primary atomization due to flash-boiling, the data is presented in the near-nozzle region only (x/D<40). The behavior of the spray at x/D = 15 is similar to one observed at x/D = 20, except that the magnitude of the spread-angle is greater owing to the bell-shaped structure of spray near to fully-flashing regime. An exciting feature is noticed at x/D = 40, where the prediction by the model coincides with the correlation curve, which is valid only in a fully flashing regime. At this downstream distance, the radial velocity of the ejected particles is insignificant, and the entrainment effect starts to dominate.

Further, the simulations are carried out for the fuels – iso-octane and acetone, and the result is shown against the experimental data at x/D=15 in Fig. 10. The experimental result is shown for the three fuels, and for each fuel, the data points correspond to the different ambient pressures. Also, the experimental results belong to the fully-flashing regime. The experimental data could not be shown for the superheat degrees corresponding to an early transition period, which is due to the paucity of spread-angle data available in the public domain, for the fuels at this range of lower superheat temperatures. As discussed above, the computational data of ethanol agree well with the available experimental results for lower superheat degrees at x/D=20 in Fig. 7(b). As visible in Fig. 10, the trend of the computed results for iso-octane and acetone is similar to that of the ethanol.

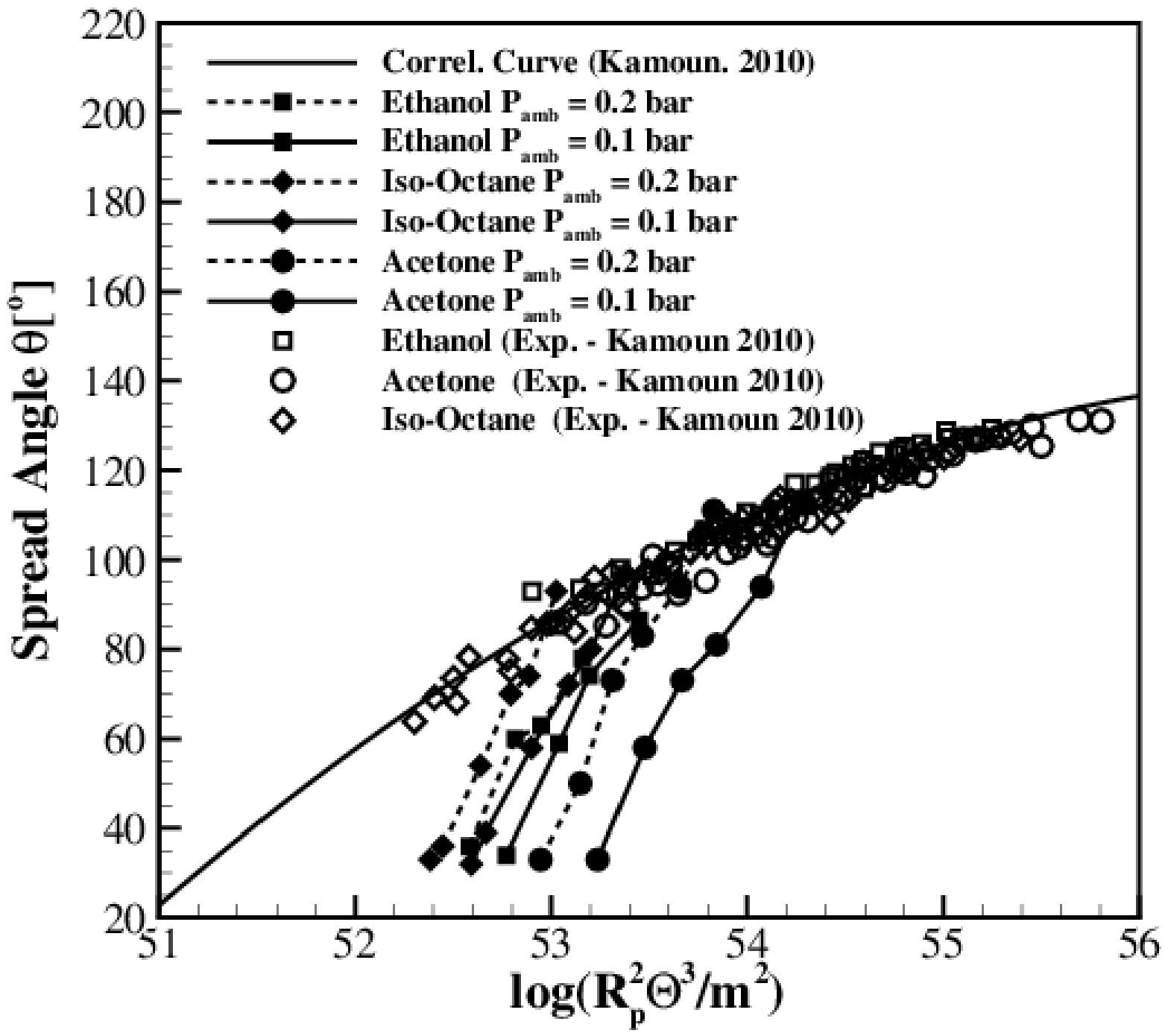

**Fig. 10.** Computed spread-angle variation of ethanol, iso-octane and acetone liquid with $\xi$ compared against the experimental results of Kamoun et al. (2010), at x/D=15.

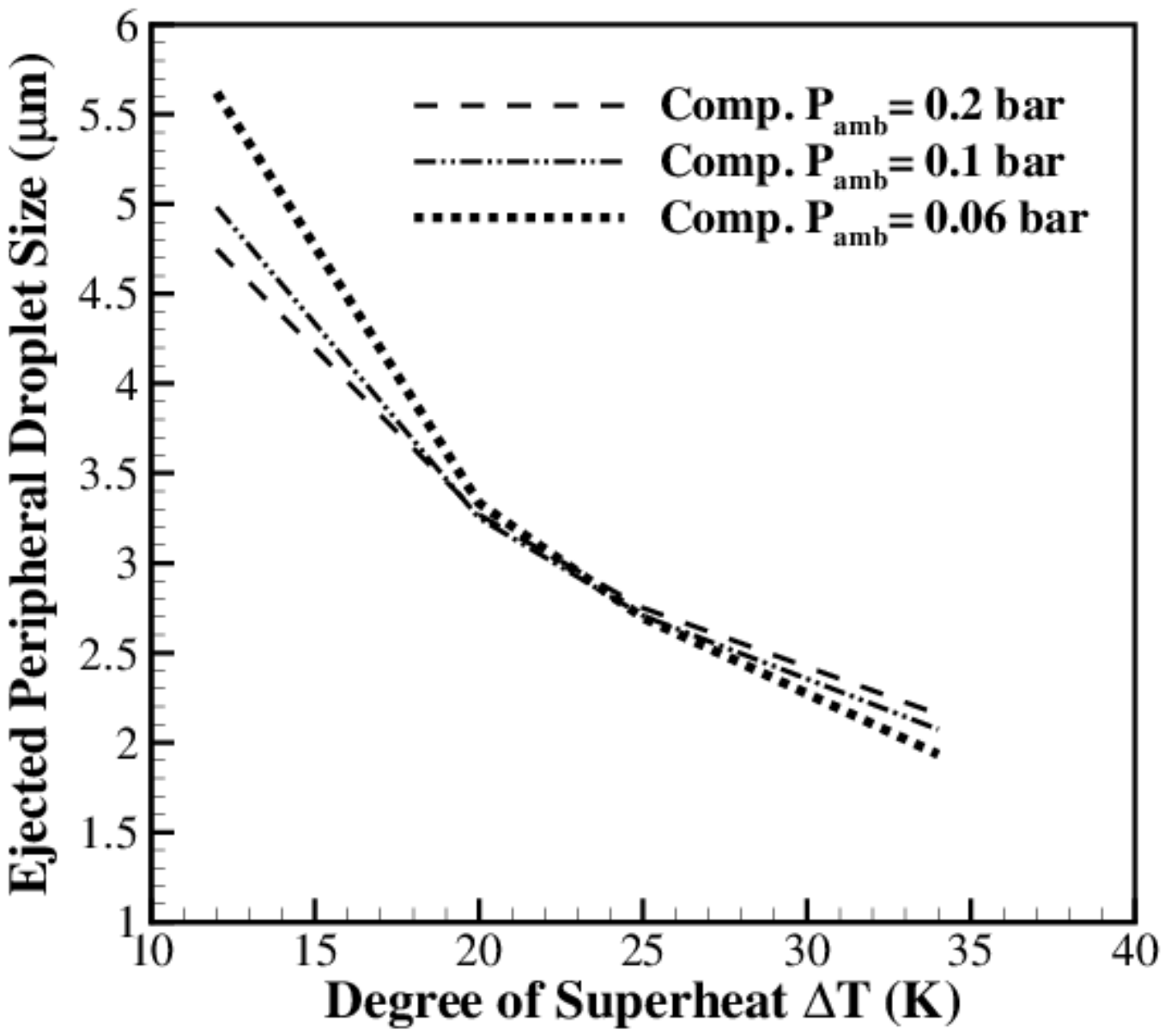

**Fig. 11.** Ejected particle size variation with pressure and degree of superheat for ethanol.

*5.3. Particle Size*

Moreover, the particle size of the ejected droplets at different superheat temperatures and pressure conditions is depicted in Fig. 11. The particle size decreases with an increase in the degree of superheat, which is in agreement with the other studies (Price et al. 2018). Finer droplet size at a higher superheat degree shows that the spray is finely atomized. Fine droplet size, along with a large number of nucleation rate at higher superheat degree, leads to faster disintegration of the liquid jet-core.

But, with an increase in $R_p$ or at lower chamber pressure, the ejected mass increases for superheat values less than 20K-25K, as shown in Fig. 11. At lower chamber pressure, the nucleation rate also increases (Kocamustafaogullari and Ishii 1983; Blinkov et al. 1993). This would again result in higher flashing, and faster disintegration of jet-core at this lower superheat degree and lower pressure values. At superheat temperatures higher than this range, the trend reverses, and the droplet size decreases with an increase in $R_p$ . This observation does not point to the slow disintegration of flashing sprays. In this range of superheat values, the nucleation rate is sufficiently higher, which not only dominates the effect of the decrease in droplet size but also results in an increase in the number of the ejected droplets, and hence, the faster disintegration of flashing sprays.

## 6. Conclusion

In the current work, we have developed a mathematical model for the specific condition of flashing sprays, where the presence of liquid jet-core causes the radial ejection of droplets. The model is validated against the experimental data (Kamoun et al. 2010) at different pressure and temperatures for three fluids. It extends the use of Dirichlet hyperboloid to replicate the droplet break-up from the liquid jet-core. Some of the assumptions in the study are meant for the simplification, and also, it may require detailed experimental investigation to make it accurately, such as the depth of the bubble is assumed constant. Overall, the method provides a unified model that takes into account several features of bubble-burst like the ejected droplet size and velocity and being more realistic concerning bubble nucleation, growth, and break-up.

The analysis of the breakup criterion of the proposed model represents the contribution of three fundamental driving forces in the bubble-burst process. It is found that the pressure-based potential energy continuously transforms into the kinetic energy, which in turn is responsible for bubble-burst.

It is observed that as a consequence of an increase in superheat degree, the radial velocity of the ejected droplets increases, which thereupon faces higher drag and thus, results in a bell-shaped structure. Similarly, the model also explains the faster disintegration of the liquid jet core with increasing temperature or decreasing chamber pressure.

## Acknowledgments

Funding has been provided by Alexander von Humboldt Foundation. Also, the authors would like to thank Prof. B. Weingard, and Dr. G Lamanna of University of Stuttgart, Germany, for sharing their experimental data for model validation and would also like to acknowledge the IITK computer center (www.iitk.ac.in/cc) for providing the resources for the computational work and data analysis.